\documentclass[twocolumn,showpacs,preprintnumbers,amsmath,amssymb,prb]{revtex4-1}
\usepackage{amsmath}
\usepackage{amssymb}
\usepackage{graphicx}
\usepackage{dcolumn} 
\usepackage{bm} 
\usepackage{longtable}
\usepackage[usenames]{color}
\usepackage[normalem]{ulem}
\usepackage[usenames]{color}
\usepackage[normalem]{ulem}

\begin{document}
\title{Tuning of the Rashba effect in Pb quantum well states \\ via a variable Schottky barrier}
\author{Bartosz Slomski$^{1,2}$}
\author{Gabriel Landolt$^{1,2}$}
\author{Gustav Bihlmayer$^{3}$}
\author{J\"urg Osterwalder$^{1}$}
\author{J. Hugo Dil$^{1,2\star}$}
\affiliation{
$^{1}$Physik-Institut, Universit\"at Z\"urich, Winterthurerstrasse 190, 
CH-8057 Z\"urich, Switzerland 
\\ 
$^{2}$ Swiss Light Source, Paul Scherrer Institut, CH-5232 Villigen, 
Switzerland\\
$^{3}$ Peter Gr\"unberg Institut and Institute for Advanced Simulation, Forschungszentrum J\"ulich and JARA, 52425 J\"ulich, Germany}
\date{\today}
\begin{abstract}
Spin-orbit interaction (SOI) in low-dimensional systems results in the fascinating property of spin-momentum locking. In a Rashba system the inversion symmetry normal to the plane of a two-dimensional (2D) electron gas is broken, generating a Fermi surface spin texture reminiscent of spin vortices of different radii. This can be exploited in a spin-based field-effect transistor (spin-FET) \cite{Datta:1990, Wolf:2001}, where the Rashba system forms a 2D channel between ferromagnetic (FM) source and drain electrodes. The electron spin precesses when propagating through the Rashba channel and spin orientations (anti)parallel to the drain give (low) high conductivity. Crucial is the possibility to tune the momentum splitting, and consequently the precession angle, through an external parameter. Here we show that this can be achieved in Pb quantum well states through the doping dependence of the Schottky barrier, opening up the possibility of a terahertz spin-FET. \\
\vspace{2cm}
$^{\star}$ Correspondence to jan-hugo.dil@psi.ch
\end{abstract}


\maketitle

The Rashba effect \cite{Bychkov:1984} has been studied quantitatively on a variety of high-Z metal containing surfaces because of the accessibility of the surface states for spin- and angle-resolved photoemission (SARPES) experiments  \cite{Dil:2009R}. Although several surfaces show very large spin splittings, most of them have metallic substrates and are thus not suitable for the design of a spin-FET due to the short-circuiting via the bulk states. Recently it was found that the surface of a topological insulator can also host Rashba-type spin split states that coexist with the non-trivial states \cite{King:2011,Zhu:2011}. However, the origin of these states is likely due to adsorption processes and an increased interlayer spacing, and they are not susceptible to a small electric field \cite{Emereev:2011arxive}. One of the main reasons why it is difficult to control the Rashba effect in surface or interface states by an external electric field is that this field should  act directly on the states themselves and thus fields in the order of $10^{13}$ Vm$^{-1}$ are required for any substantial change \cite{Bihlmayer:2006}. For semiconductor heterostructures, on the other hand, the applied gate voltage indirectly affects the states via the symmetry of the confinement, and much lower voltages are required for a measurable change \cite{Nitta:1997,Studer:2009}. 

Here we propose that quantum well states (QWS) in an ultra-thin metal film grown on a semiconductor may offer an alternative way to realize a system where the Rashba effect can be manipulated by a small gate voltage. Measured momentum splittings are larger than in semiconductor heterostructures, indicating that much shorter channel lengths may here be functional for spin-FET-type devices. We focus on Pb films on Si(111) samples with different doping levels as a model system, because (i) Pb is a high-Z element (Z = 82) with large atomic SOI, (ii) smooth crystalline films can be grown on Si(111) with chemically sharp and well-defined interfaces \cite{Upton:2004,Ricci:2004}, and (iii) the electric field in the depletion layer due to the ionized donors and acceptors can be varied by a gate voltage.

Recently it was shown that QWS in thin Pb films on n-type Si(111) show a Rashba-type spin splitting which builds up throughout the whole metal layer along the growth direction as a result of competing effects between the metal-substrate and metal-vacuum interfaces \cite{Dil:2008}. This is indicated by the absence of a strong dependence of the strength of the Rashba effect on the film thickness, and by the reversed spin helicity as compared to the surface state on Au(111) \cite{Hoesch:2004}. This is in contrast to other ultra-thin film systems where the spin splitting of QWS is induced by a high-Z metal substrate \cite{Varykhalov:2008b,Rybkin:2010}, or by hybridization gaps \cite{He:2008}. Later the sensitivity of the Rashba parameter to the film-substrate interface was determined in Pb QWS: replacing the Pb wetting layer by a Bi layer reduces the Rashba parameter by 50~\%,  whereas QWS in Pb films grown on a Ag reconstructed Si substrate show no measurable Rashba-type spin splitting \cite{Slomski:2011B}. As will be explained below, these effects can be understood in terms of the local asymmetry of the QWS wave function with respect to Pb nuclei, which in turn is related to the particular confinement conditions. Here we will show that a similar effect can be produced by changing the donor concentration of the substrate which subsequently alters the Schottky barrier (SB) of the metal-semiconductor junction.
\begin{figure}[htb]
\begin{center}
\includegraphics[width=0.48\textwidth]{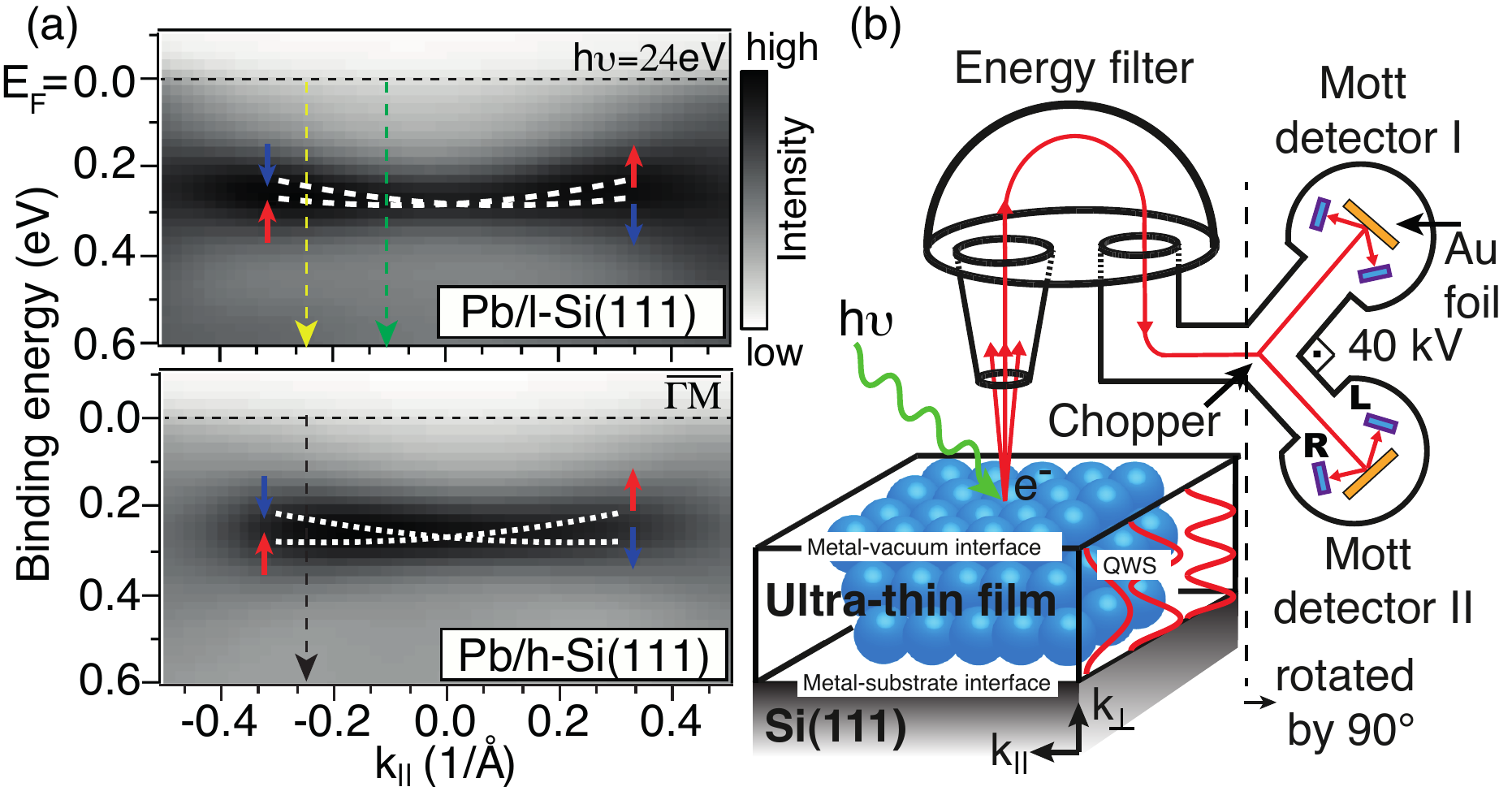}
\caption{(color online) (a) Band dispersion of a QWS in 8 ML thick Pb films on (upper panel) lightly n-doped and (lower panel) heavily n-doped Si(111) substrates as measured by ARPES. (b) Schematic drawing of our SARPES experiment.} 
\label{Fig1}
\end{center}
\end{figure}

Figure \ref{Fig1}(a) displays ARPES data from QWS formed in an N = 8 monolayers (ML) thick Pb film deposited on lightly (upper panel) and heavily (lower panel) n-doped Si substrates [henceforth Pb/h-Si(111) and Pb/l-Si(111)]. The band dispersions are very similar to those obtained for Pb films grown on a moderately n-doped Si(111) substrate [hereafter Pb/m-Si(111)]\cite{Mans:2002,Upton:2005,Dil:2006}. Both band dispersions feature an anomalously high effective mass ($>$10 m$_{e}$), which is due to an increased in-plane lattice constant \cite{Slomski:2011}. 
The films on both samples  were measured with the COPHEE spectrometer \cite{Hoesch:2002} equipped with two Mott detectors, see Fig. \ref{Fig1}(b), and were prepared by the same procedure. A change of the spin splitting is therefore directly related to the influence of the donor concentration in the Si substrate, because the metal-vacuum boundaries are the same. 
\begin{figure}[htb]
\begin{center}
\includegraphics[width=0.48\textwidth]{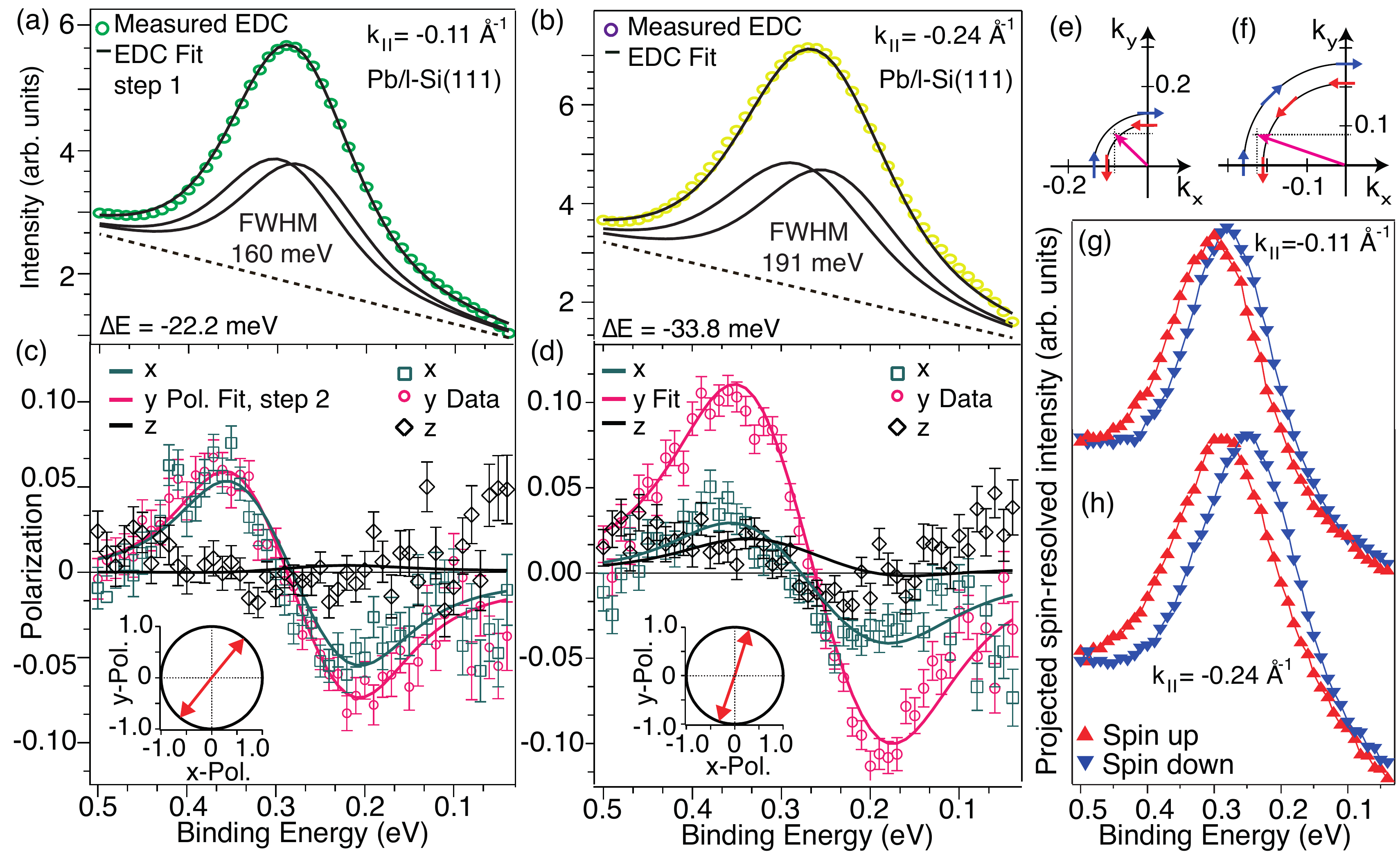}
\caption{(color online) Measured spin-integrated EDCs for Pb/l-Si(111) (a) at k$_{\parallel}$ = -0.11 \AA$^{-1}$ and at (b) k$_{\parallel}$ = -0.24 \AA$^{-1}$ with Voigt profiles obtained from the self-consistent two-step fitting routine. (c, d) Corresponding spin polarization data and fits of the x, y and z components. Insets show the in-plane polarization vectors obtained from the two-step fit. (e, f) Parts of constant energy surfaces illustrating the measurement positions of the EDCs in reciprocal space . (g, h) Spin-resolved intensity profiles projected on the quantization axis for the two measured momenta.} 
\label{Fig2}
\end{center}
\end{figure}

Figure \ref{Fig2} shows SARPES data from a Pb QWS grown on the lightly n-doped Si(111) measured for two energy distribution curves (EDCs) at (a) k$_{\parallel} = -0.11$ \AA$^{-1}$  and (b) k$_{\parallel}=-0.24$ \AA$^{-1}$, see also dashed arrows in the upper panel of Fig. \ref{Fig1}(a) which indicate the positions where the spin-resolved spectra were recorded. For the measurements at both momenta displayed in Figs. \ref{Fig2}(c, d) the spin polarization data show typical up-down excursions in the tangential spin polarization components as expected for a Rashba system. 
A quantitative analysis of the spin-resolved data using a two-step fitting routine \cite{Meier:2009NJP} reveals a larger spin splitting between the spin-up ($\uparrow$) and spin-down ($\downarrow$) bands ($\Delta E := E_b^{\downarrow}(k_{\parallel})-E_b^{\uparrow}(k_{\parallel})$) for the state with higher momentum. The corresponding fits are shown in Figs. \ref{Fig2}(a, c) for k$_{\parallel} = -0.11$ \AA$^{-1}$ and in (b, d) for k$_{\parallel} = -0.24$ \AA$^{-1}$. We obtain a spin splitting of $\Delta E = -(22.2 \pm 1.4)$ meV and $-(33.8 \pm 1.4)$ meV, respectively, with fully polarized bands. 
The increased spin splitting with increasing momentum is in full agreement with the Rashba model \cite{Bychkov:1984} where $\Delta E \propto$ k$_{\parallel}$, and also seen in Figs. \ref{Fig2}(g) and (h), which display raw data as spin-resolved EDCs projected onto the quantization axis at both momenta, calculated as $I^{\uparrow, \downarrow}_{\text{tan}} = \frac{1}{2}I_{\text{tot}}(1 \pm P_{\text{tan}})$ with $P_{\text{tan}} = \text{sign}(P_y)\sqrt{P_x^2 + P_y^2}$. Figures \ref{Fig2}(e) and (f) indicate the precise parallel momentum positions where the two EDCs were measured. For $k_{\parallel} = -0.11$ \AA$^{-1}$ the position was $(k_{x}, k_{y}) = (-0.086, 0.072)$ \AA$^{-1}$ (Fig. \ref{Fig2}(e)),  for $k_{\parallel} = -0.24$ \AA$^{-1}$ it was $(k_{x}, k_{y}) = (-0.227, 0.072)$ \AA$^{-1}$ (Fig. \ref{Fig2}(f)). The strict Rashba-type spin-momentum locking requires the spins to be quantized in the direction perpendicular to the momenta, and the observed polarization along the x-direction is thus a consequence of the non-zero k$_{y}$ component.

\begin{figure}[htb]
\begin{center}
\includegraphics[width=0.45\textwidth]{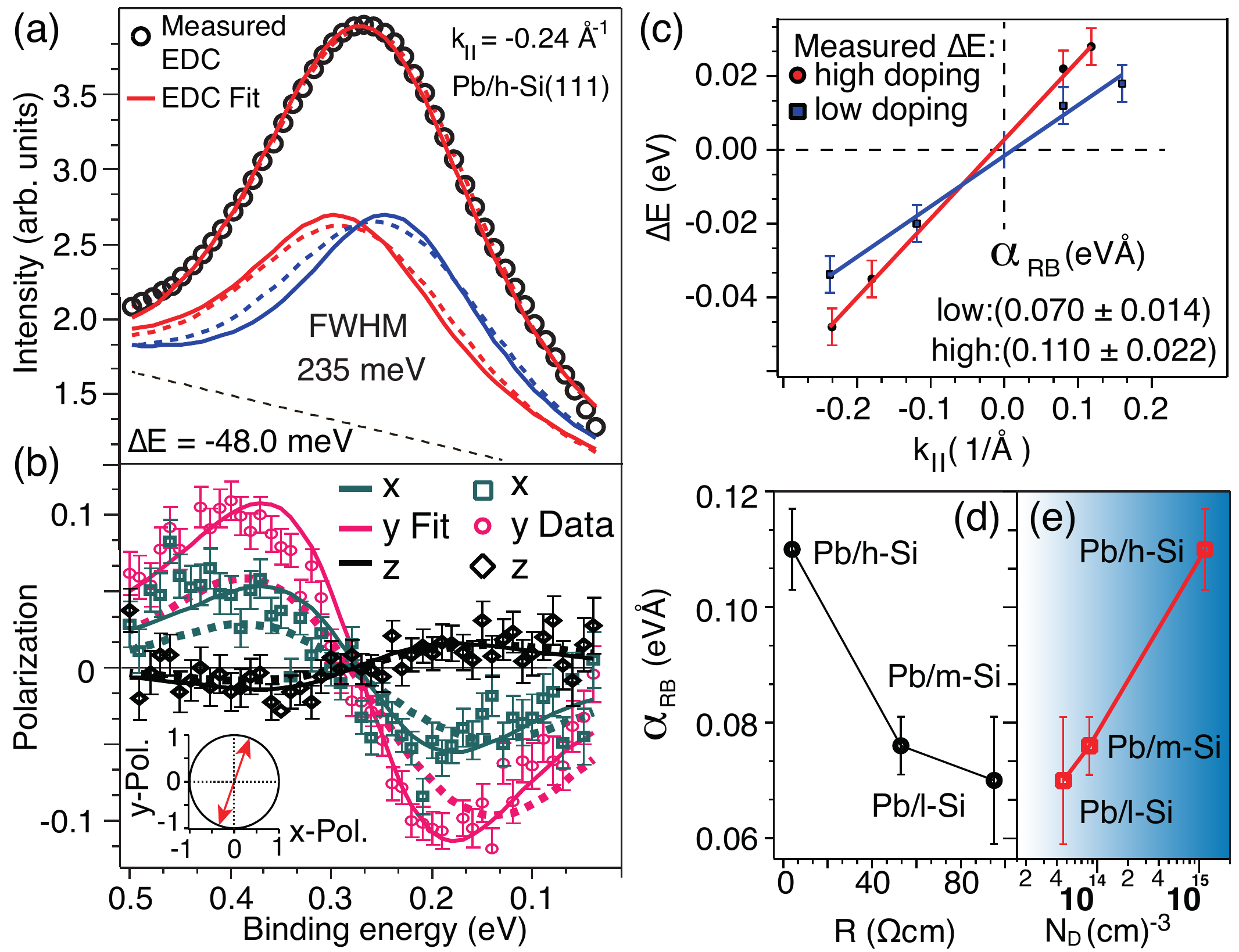}
\caption{(color online) Measured spin-integrated EDC for Pb/h-Si(111) at k$_{\parallel}$ = -0.24 \AA$^{-1}$ (a) with Voigt intensity profiles obtained from the two-step fit. (b) Corresponding polarization data and fits of the x, y and z components. The dashed lines are fits performed with $\Delta$E = -33.8 meV. (c) Measured k-dependent energy splittings and linear fits (lines) to obtain $\alpha_{RB}$ for Pb/h-Si(111) and Pb/l-Si(111), respectively. (d) Measured $\alpha_{RB}$ of the differently n-doped substrates vs. resistance and (e) donor concentration.} 
\label{Fig3}
\end{center}
\end{figure}

Figures \ref{Fig3}(a-b) show SARPES data at k$_{\parallel} = -0.24$ \AA$^{-1}$ for a Pb film of the same thickness prepared on the heavily n-doped Si(111). The best self-consistent fit to the intensity and spin polarization data is here achieved with $\Delta E = -(48.0 \pm 1.8)$ meV. A direct comparison of both systems at k$_{\parallel} = -0.24$ \AA$^{-1}$ reveals a larger energy splitting on the heavily n-doped substrate by almost 14 meV, which is well beyond the combined accuracy of 2 meV  of the measurement and the fitting routine \cite{Meier:2009NJP}. 
To illustrate that the larger $\Delta E$ of the QWS in Pb/h-Si(111) is not an artifact from our analysis procedure, we have performed a fit by keeping $\Delta E$ equal to the value found in Pb/l-Si(111) (dashed lines in Fig. \ref{Fig3}(a, b)). Although the total intensity fit is reasonably good, the fit to the polarization data obviously fails. 

From these and further data sets measured at different $k_{\parallel}$ we can deduce the Rashba constants for both doping levels by plotting $\Delta E$, obtained from the fitting, vs. $k_{\parallel}$ (Fig. \ref{Fig3}(c)). Values for $\alpha_{\text{RB}}$ obtained by using $\alpha_{\text{RB}}~=~
1/2\cdot(\text{d}\Delta E)/(\text{d}k_{\parallel})$ are $\alpha_{\text{RB,h}}  = (0.11 \pm 0.010)$ eV\AA $\hspace{1mm}$ for Pb/h-Si(111), $\alpha_{\text{RB,m}} = (0.076 \pm 0.005)$ eV\AA $\hspace{1mm}$ for Pb/m-Si(111)\cite{Slomski:2011B}, and $\alpha_{\text{RB,l}} = (0.070 \pm 0.007)$ eV\AA $\hspace{1mm}$ for Pb/l-Si(111).

We now discuss the mechanism leading to the decrease of the Rashba constant in Pb QWS with increasing substrate resistance, as shown in Fig. \ref{Fig3}(d), or equivalently the increase of $\alpha_{RB}$ with donor concentration (Fig. \ref{Fig3}(e)). 
As demonstrated in Ref. \cite{Slomski:2011B} by the comparison of the Rashba effect in Pb QWS on a Pb and Bi terminated Si surface, a changed Rashba constant results from a modified charge density distribution in the Pb film which alters the local asymmetric features of the QWS wave function around the Pb cores. 
According to the phase accumulation model the charge density distribution of a QWS is controlled by the phase shifts at the metal-substrate and metal-vacuum interfaces \cite{Chiang:2000}. Because the phase shift at the vacuum side is the same for all three investigated systems, we focus in the following on the 
metal-substrate interface with its phase shift given by $\Phi_S \propto \sqrt{E-E_0}\Theta(E-E_0)$ \cite{Ricci:2004} where $E$ is the energy of a QWS and $E_0$ is the valence band edge of Si at $\overline{\Gamma}$ and $\Theta$ is the Heaviside function. States with energies inside the gap of a substrate are truly confined, while states outside the gap can couple to substrate states leading to quantum well resonances with a considerable fraction of charge spilling into the substrate. In between these two cases the confinement, i.e. the degree of localization, changes as a function of $E_0$. The goal of the following discussion is to show that the energetic distance between the QWS and $E_0$ is sensitive to the Schottky barrier ($\Phi_{SB}^{n}$) and hence to the donor concentration because both quantities are related via: $\Phi_{n}^{SB} + E_0 = E_g$ where $E_g$ is the energy gap of Si(111).

Our explanation is based on the interface dipole model for Schottky barrier formation, which takes the interaction of the metal and the substrate into account \cite{Tung:2000}. In this model the formation of polar Pb--Si bonds at the interface results in a charge transfer and the establishment of an interface dipole that directly contributes to the SB \cite{Tung:2000}: 
\begin{eqnarray}
\Phi_{n}^{SB} = \Phi_M - \chi_S + eV_{int}. 
\end{eqnarray}
Here $\Phi_M$ is the metal work function, $\chi_S$ is the electron affinity of the semiconductor, and $V_{\text{int}}$ is the dipole induced voltage drop at the interface. We will argue that the size of $V_{\text{int}}$ is not only determined by the interface chemistry, but also by the doping concentration and the type of dopant (donor or acceptor).

\begin{figure*}[htb]
\begin{center}
\includegraphics[width=0.9\textwidth]{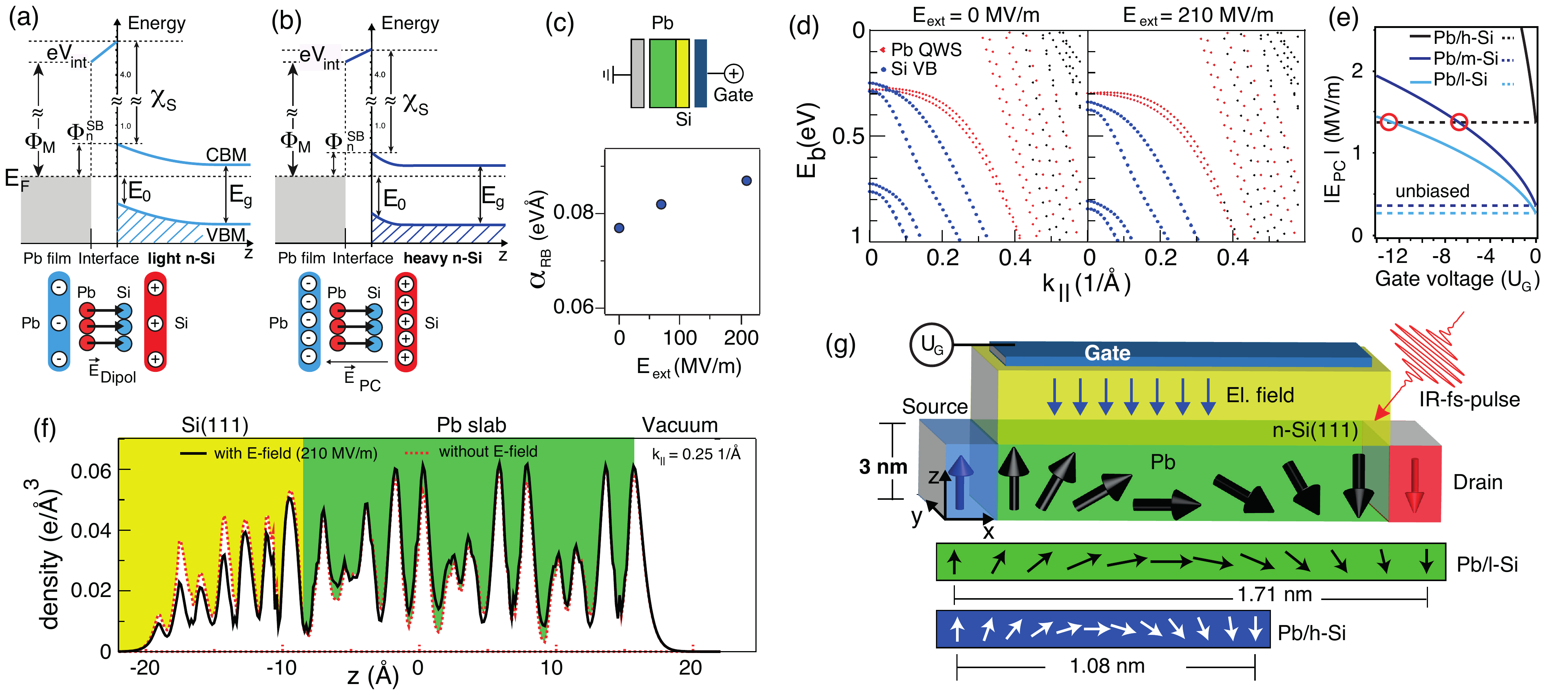}
\caption{(color online) (a,b) Energy diagrams of Pb/Si within the interface dipole model (not to scale) as a function of $N_D$ (upper panels), and capacitor-dipole model (lower panels). (c) Pb/Si in a capacitor to illustrate how $E_{\text{ext}}$ is applied to the system (upper panel). Calculated Rashba constant of Pb/Si as a function of applied external E-field (lower panel).  (d) Influence of $E_{\text{ext}}$ on the energetic position of the Si valence band edge. (e) Electric field in the parallel plate capacitor as a function of (reversed) gate voltage. (f) Charge density plots of QWS in Pb/Si with (solid line) and without (dashed line) $E_{\text{ext}}$ at $k_{\parallel}$ = 0.25 \AA$^{-1}$. (g) Sketch of a spin-FET device operated with a gate voltage ($U_G$) or by excitation with femto-second IR pulses (upper panel). Spin precession of $\pi$ for electrons propagating along the x-direction in Pb films on l-Si(111) and h-Si(111) (lower panels).} 
\label{Fig4}
\end{center}
\end{figure*}

The upper panel of Fig. \ref{Fig4}(a) shows the energy band diagram of Pb/Si deduced from measuring the peak widths of QWSs and QWRs as suggested in Ref. \cite{Ricci:2004}, and from core-level shifts of the Si 2p spectrum in the interface region that revealed an accumulation of negative charge at Si atoms when the Pb-Si interface is formed. This is in-line with the predictions of the polarity of the Pb-Si bond using the Miedema electronegativity scale and with the study of Si 2p core-level shifts of Pb on n- and p-type Si that both revealed a shift toward lower binding energies \cite{LeLay:1990}. 
According to Koopman's initial state model this means that the electric field generated by the interface dipole ($\mathbf{E}_{\text{Pb-Si}}$) points from Pb to Si independent of the type of dopant. The lower panel of Fig. \ref{Fig4}(a) illustrates our model which explains the influence of the donor concentration on the band edge of Si(111).
It consists of the aforementioned interface dipole placed in a parallel plate capacitor (PC) that generates an electric field of strength \cite{Brillson:1993} $|\mathbf{E}_{\text{PC}}| =\sqrt{2eN_D(U_0 + U_G)/\epsilon\epsilon_0}$ at the interface due to the space charge in the depletion layer  and hence depends on the density of ionized donors ($N_D$). 
$U_0$ is the built-in potential across the Schottky barrier, $U_G$ is the gate voltage (for the moment $U_G$ = 0 V) and $\epsilon = 11.9$ the dielectric constant of Si. 
For Pb in contact with n-type (p-type) Si the polarity of the capacitor is such that the positive (negative) space charge is on the semiconductor side, while the negative (positive) charge of equal size is balanced on the Pb side. Hence the direction of the electric field of the dipole and of the capacitor are \emph{anti-parallel} (parallel) to each other for n-type (p-type) Si. Now, increasing the donor concentration increases the electric field in the capacitor and lowers $V_{int}$ due to the screening of the interface dipole. Consequently, the SB decreases and the band edge of Si shifts to higher binding energies. 

Our finding of a reduction of the SB with increasing $N_D$ is consistent with previous results \cite{Kang:2012, Zhang:2010}. As already pointed out the increased energetic distance between the QWS and $E_0$ influences the confinement of the QWS via the metal-substrate phase shift such that the QWS is more strongly localized in the Pb film, i.e. less charge spills into the substrate. This affects the local charge densities of the QWS wave function with respect to the Pb cores which determine the spin splitting \cite{Slomski:2011B}.

We now show how the charge density distribution in the Pb film changes as a function of the energetic position of the valence band edge using DFT calculations. These calculations were performed on a 10 ML Pb film on Si(111) with the in-plane lattice constant of the Pb film commensurate with that of Si(111) and with Pb/Si placed into a parallel plate capacitor producing external electric fields ($E_{\text{ext}}$) of various strengths with the positive bias at the Si side, see upper panel of Fig. \ref{Fig4}(c). The DFT calculations reveal that increasing $E_{\text{ext}}$ from 0 to 210 MVm$^{-1}$ shifts the Si valence band edge by $\approx$ 90 meV to higher binding energies, see Fig. \ref{Fig4}(d), and simultaneously the calculated Rashba constant increases by $\approx$ 12 \% , see lower panel of Fig. \ref{Fig4}(c). Notice that here $E_0$ shifts in the same direction as in the interface dipole model when the donor concentration is increased. As already speculated the increased Rashba constant is a result of an increased localization of the state in the Pb film, which is apparent from Fig. \ref{Fig4}(f) where we show charge density distributions producing a $\alpha_{RB}$ = 0.078 eV\AA\ at $E_{\text{ext}}$ = 0 MVm$^{-1}$ and 0.087 eV\AA\ at 210 MVm$^{-1}$, respectively. This interpretation is also consistent with our results obtained from Pb QWS on heavily p-doped Si(111) (acceptor concentration $N_A = 1.18 \cdot 10^{16}$ cm$^{-3}$) where we find the smallest Rashba constant among the studied systems of $\alpha_{RB}^{Pb/p-Si} = (0.061 \pm  0.004)$ eV\AA, see SOM.

Having shown the sensitivity of the Rashba constant in Pb QWS to the donor concentration we now discuss the possibility to use Pb/Si as a candidate for future spintronic applications. Assuming ballistic spin transport through the Rashba channel we deduce the length (L) at which the injected spin - a coherent superposition of the two orthogonal Rashba spinors - precess by an angle of $\pi$ while propagating along $x$, using L = $\pi/(2k_0)$ where $2k_{0} = 2m^{\star}\alpha_{RB}/\hbar^2$ is the characteristic Rashba-type momentum splitting deduced from the experiment. For electrons in Pb/l-Si(111) we obtain L$_{l}$ = 1.71 nm from k$_{0,l}$ = 0.092 \AA$^{-1}$, and in Pb/h-Si(111) L$_{h}$ = 1.08 nm from k$_{0,h}$ = 0.146 \AA$^{-1}$, see lower panels of Fig. \ref{Fig4}(g). For a spin-FET made of Pb/l-Si(111) with a fixed channel length of 1.08 nm this means that changing the voltage drop at the interface by increasing the dipole screening via $|\mathbf{E}_{\text{PC}}| \propto \sqrt{U_0+U_G}$ it is possible to go from a low ($\Delta \varphi \approx \pi/2$) to a high conducting state ($\Delta \varphi = \pi$)  by ramping $U_G$ from 0 to $-12.8$ V as simulated in Fig. \ref{Fig4}(e).

An alternative way to operate a spin-FET device may also be achieved by optical means -- thereby combining spin-based with electro-optical technologies. Recently it has been demonstrated that it is possible to change the SB and correspondingly $E_0$ in Pb/Si using ultra-short light pulses \cite{Rettig:2012}. Now, consider a spin-FET in the non-conducting state which can be optically driven. Illuminating the device with fs-pulses leads to a shift of $E_0$ such that $k_{0}$ changes and the conducting state is reached. For high-speed applications, the FET should respond quickly to variations of the trigger pulse. In Pb/Si a complete built-up of the $E_0$ shift is reached within 100 fs and equilibrium is recovered after 600 fs \cite{Rettig:2012} - the non-conducting state. This yields a switching frequency in the \emph{terahertz} (THz) regime. 

Our results demonstrate how the Rashba-type spin splitting of QWS in ultra-thin Pb films can be tuned effectively via the donor concentration of the n-Si(111) substrate. We conjecture that the energetic position of the Si valence band edge is the decisive factor for this effect. This opens up the possibility to fine-tune the Rashba-type momentum splitting in metallic QWS via a gate voltage or by excitation with ultra-short laser pulses. 

We thank C. Hess, F. Dubi, M. Kropf and S. Stutz for technical support and G. Salis for discussion of the results. This work was supported by the Swiss National Foundation.

\subsubsection{Methods}
The ultra-thin Pb films were prepared in three steps consisting of (i) cleaning of the substrate by flashing above T $>$ 1300 K to obtain the Si(111)-(7$\times$7) reconstruction, (ii) deposition of $\approx$ 3 ML of Pb from a Knudsen cell and subsequent annealing at 600 K for 60 sec. to form the dense $(\sqrt{3}\times \sqrt{3})$R30$^{\circ}(\alpha)$ surface reconstruction \cite{Weitering:1992}, and (iii) deposition of thin crystalline Pb films at T~=~80 K. The films on both substrates were prepared by depositing the same amount of Pb with a rate of $\approx$ 0.3 ML/min.
The room temperature substrate resistance $R$ and donor concentrations N$_{D}$ are for  Pb/h-Si(111): R = 4 $\Omega$cm,  N$_{D}$ = 1.14 $\cdot 10^{15}$ cm$^{-3}$, for Pb/m-Si(111): R = 53 $\Omega$cm,  N$_{D}$ = 8.38 $\cdot 10^{13}$ cm$^{-3}$ and for Pb/l-Si(111): R = 95 $\Omega$cm,  N$_{D}$ = 4.65 $\cdot 10^{13}$ cm$^{-3}$. 

All measurements were performed with the COPHEE end-station at the Swiss Light Source of the Paul-Scherrer-Institute \cite{Hoesch:2002} at a base pressure below 3$\times$10$^{-10}$ mbar. The sample temperature during the measurements was 80 K. The energy and angular resolution of our spin-resolved (spin-integrated) ARPES measurements were set to 100 (25) meV and $\pm$ 0.75$^{\circ}$ (0.5$^{\circ}$) in respect of the low efficiency of Mott-scattering. In all measurements a photon energy of 24 eV was used because of the optimal photoemission cross-section for Pb QWS \cite{Dil:2004}. 

The ARPES spectrometer is equipped with two orthogonally mounted Mott detectors operated at an acceleration voltage of 40 kV \cite{Petrov:2003} (see Fig. \ref{Fig1}(b)). Each Mott detector measures asymmetries of the photoelectron current backscattered elastically off a polycrystalline gold foil, defined as $A_{i} = (I_{R} - I_{L})/(I_{R} + I_{L})$, with $I$ being the intensity measured in either right (R) or left (L) detector. The orthogonal arrangement of the two Mott detectors allows to measure asymmetries in A$_{\tilde x}$, A$_{\tilde y}$ and A$_{\tilde z}$ directions of the Mott coordinate system ($\tilde x, \tilde y, \tilde z$). Three-dimensional spin polarization data is then obtained by dividing asymmetries by the Sherman function (S = 0.068), and by applying a rotation matrix (T) in order to transform the spin polarization components from the Mott reference frame into the sample coordinate system: $P_{i}(E_{b}) = T(S^{-1}A_{\tilde i}(E_{b}))$ with i $\in$ (x,y,z).

We employed density functional theory in the local density approximation \cite{Perdev:1981},
using the full-potential linearized augmented plane-wave method in thin film
geometry \cite{Krakauer:1979, Flapw}. This allows us to expose the structure, consisting of a ten
layer Pb film placed commensurately on six layers of Si(111) substrate with a
H-terminated backside, to an external electric field. This field is generated
by effective condenser plates put in the vacuum at a distance of 5.3 \AA above
and below the film structure. To simulate the Si-doping, we charge only the
condenser plate below the Si and add the screening charge to the Si substrate.

\begin{footnotesize}

\end{footnotesize}

\section{Supplemental Material}
In this supplementary online material we present data and analysis of the Rashba-type spin splitting in Pb QWS in an ultra-thin Pb film of 8 ML thickness prepared on heavily p-doped Si(111). We find that the Rashba parameter is the smallest among the here studied systems which is in-line with the presented interface dipole model and the corresponding screening of the interface dipole. We will also show that the influence of the doping concentration on the spin splitting in Pb QWS can neither be explained by a semi-classical approach via the penetration depth of the QWS wave function into the barrier at the metal-substrate interface nor by the bare Rashba model. 
\subsection{Rashba effect in Pb QWS on p-type Si(111)}
In our study we have also measured spin splittings of QWS prepared on heavily p-doped Si(111) [hereafter Pb/p-Si]. Our results are summarized in Fig. \ref{Fig1}. Figure  \ref{Fig1}(a) displays the electronic structure of 8 ML Pb on p-type Si with a bulk resistivity of 1.25 $\Omega$cm and an acceptor concentration of $N_A = 1.18 \cdot 10^{16}$ cm$^{-3}$. The band dispersion is similarly flat as found for Pb on n-type Si. 

\begin{figure}[htb]
\begin{center}
\includegraphics[width=0.48\textwidth]{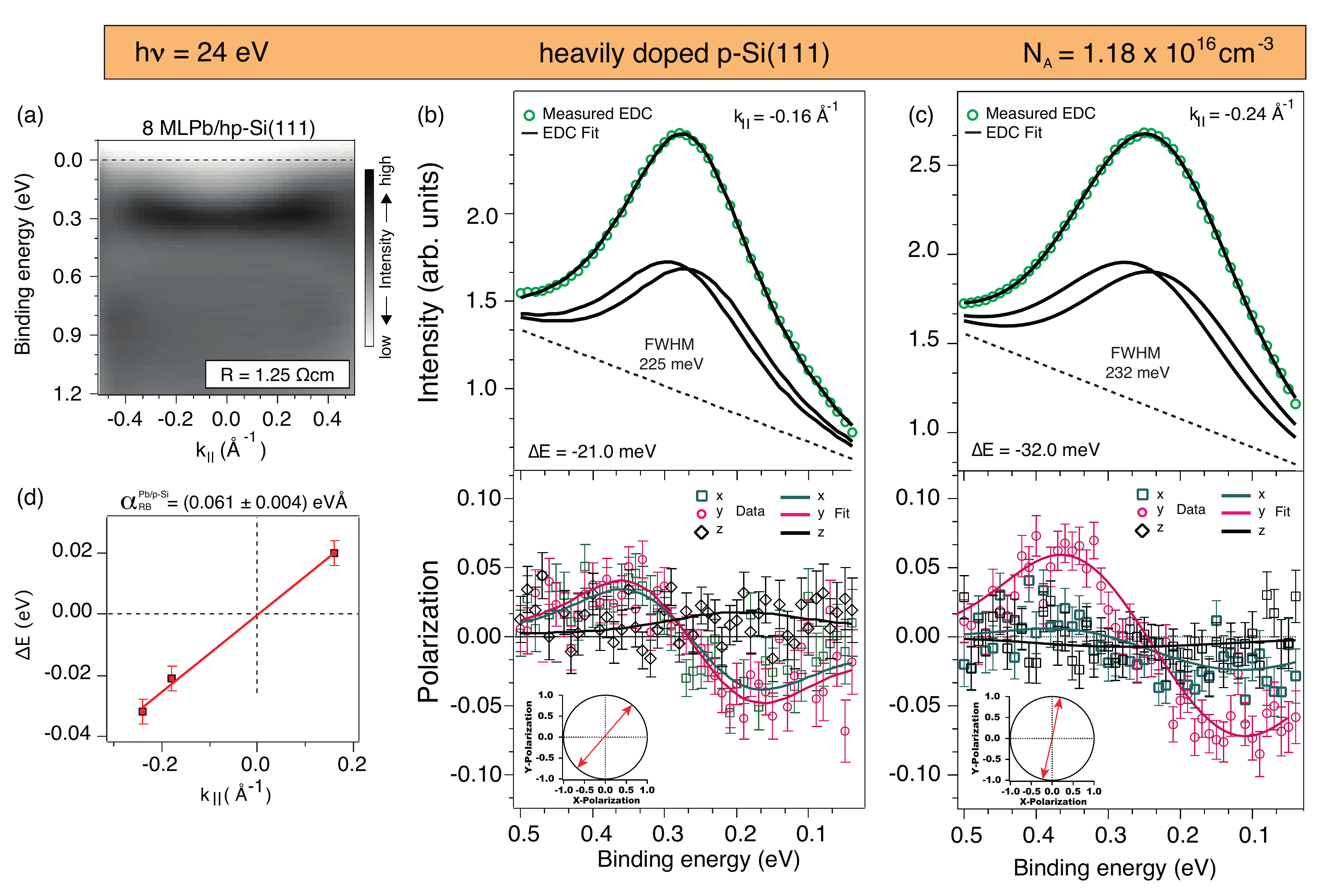}
\caption{(a) Band dispersion of a QWS in 8 ML thick Pb film on heavily p-doped. Measured spin-integrated EDCs for Pb/p-Si(111) (b) at k$_{\parallel}$ = -0.16 \AA$^{-1}$ and at (c) k$_{\parallel}$ = -0.24 \AA$^{-1}$ with Voigt profiles obtained from the self-consistent two-step fitting routine. The lower panels show the corresponding spin polarization data and fits in x, y and z-direction. Insets show the in-plane polarization vectors obtained from the two-step fit. (d) Measured k-dependent energy splittings and linear fit (line) to obtain $\alpha_{RB}$ for Pb/p-Si(111).}  
\label{Fig1}
\end{center}
\end{figure}

The upper panels of Figs. \ref{Fig1}(b) and (c) show the measured spin-integrated spectra and the lower panels spin polarization data together with fits that were taken at two different in-plane momenta, i.e. (a) at $k_{\parallel} = -0.16$ \AA$^{-1}$ and (b) at $k_{\parallel} = -0.24$ \AA$^{-1}$, respectively. The measured spin polarization data reflect Rashba-type spin splitting which manifests itself in a tangential spin alignment to the energy contour (not shown) and an increase of spin splitting with increasing momentum. The two-step fitting routine gives a spin splitting of $\Delta E = - 21.0$ meV at $k_{\parallel} = -0.16$ \AA$^{-1}$ and $\Delta E = -32.0$ meV at $k_{\parallel} = -0.24$ \AA$^{-1}$.
A direct comparison of the measured spin splittings at $k_{\parallel} = -0.24$ \AA$^{-1}$ reveals the smallest spin splitting for Pb QWS on the heavily p-doped Si substrate [cf. Pb/l-Si: $-33.8$ meV, Pb/h-Si: $-48.0$ meV].
Correspondingly, when analyzing the spin splittings as a function of the in-plane momentum we find the smallest Rashba parameter in Pb/p-Si being $\alpha_{RB}^{Pb/p-Si} = (0.061 \pm 0.004)$ eV\AA, see Fig. \ref{Fig1}(d).

\begin{figure*}[htb]
\begin{center}
\includegraphics[width=0.65\textwidth]{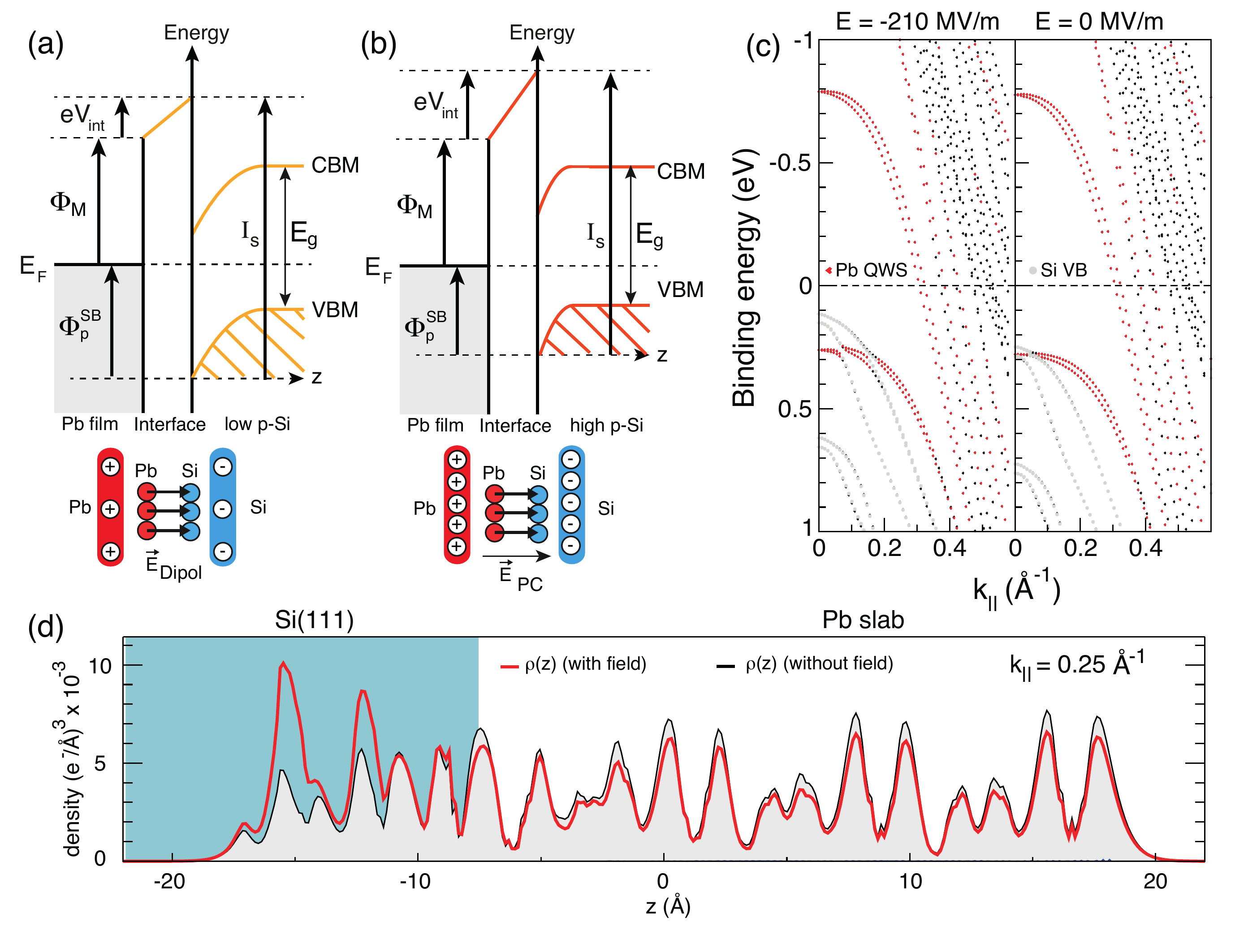}
\caption{(a,b) Energy diagram of Pb on p-type Si within the interface dipole model (not to scale) as a function of acceptor concentration (upper panels), and capacitor-dipole model (lower panels). Notice that the polarity of the capacitor is inverted with respect  to Pb on n-type Si. (c) Calculated electronic structure of 10 ML Pb on Si with (left panel) $E_{\text{ext}}$ = -210 MVm$^{-1}$ and (right panel) $E_{\text{ext}}$ = 0 MVm$^{-1}$. (c) Corresponding charge density plots of the Pb QWS as a function of the electric field.}  
\label{Fig2SOM}
\end{center}
\end{figure*}

We now show that the reduction of the Rashba parameter is well explained by the interface dipole model although, at a first glance, this finding seems to be in conflict with the found dependence of $\alpha_{RB}$ in Pb QWS on n-type Si that increased with increasing donor concentration. 

In Figs. \ref{Fig2SOM} (a, b) we draw the schematic energy band diagram for Pb in contact with p-type Si for the case of (a) low and (b) high acceptor concentration. In the framework of the interface dipole model the Schottky barrier is calculated via \cite{Tung:2000}
\begin{eqnarray}
\phi_p^{SB} = I_S - \Phi_M + eV_{int}
\end{eqnarray}
where $I_S$ is the ionization energy of the semiconductor, $\Phi_M$ is the work function of the metal and $V_{int}$ the dipole induced voltage drop. 

According to Figs. \ref{Fig2SOM}(a, b) the valence band edge of Si(111) equals the SB, i.e. $E_0 = \phi_p^{SB}$. As already mentioned in the main text, the polarity of the parallel plate capacitor in Pb/p-Si is such that the electric field generated from the ionized (negative) acceptors points from the Pb toward the Si side and is therefore parallel to the electric field generated by the interface dipole. The parallel alignment of both electric fields results in an inverted dependence of the screening of the dipole upon increasing the acceptor concentration; increasing the acceptor concentration increases the electric field in the capacitor and due to the parallel alignment the voltage drop at the interface increases. Consequently the SB decreases ($I_S$ and $\phi_M$ are constant) and so does the energetic distance between the band edge and the QWS. The decreased energetic distance alters the phase shift at the metal-substrate interface such that the QWS becomes more delocalized in the Pb film. This lowers the Rashba parameter because less charge density contributes to the local spin splittings at the Pb cores. 

Our explanation is well reproduced by DFT calculations. Figure \ref{Fig2SOM}(c) shows the band structure of 10 ML Pb on Si(111) with an applied external electric field of $-210$ MVm$^{-1}$ (left panel) and with $E_{ext}$ = 0 MVm$^{-1}$ (right panel). The electric field shifts the band edge of Si by roughly 120 meV toward lower binding energies, i.e. in the same direction when the acceptor concentration increases. The influence of this band shift on the charge density distribution of the QWS is displayed in Fig. \ref{Fig2SOM}(d). The charge density in the case of $E_{ext} \neq$ 0 MVm$^{-1}$ spills to a larger amount into the Si substrate.

\begin{figure}[htb]
\begin{center}
\includegraphics[width=0.48\textwidth]{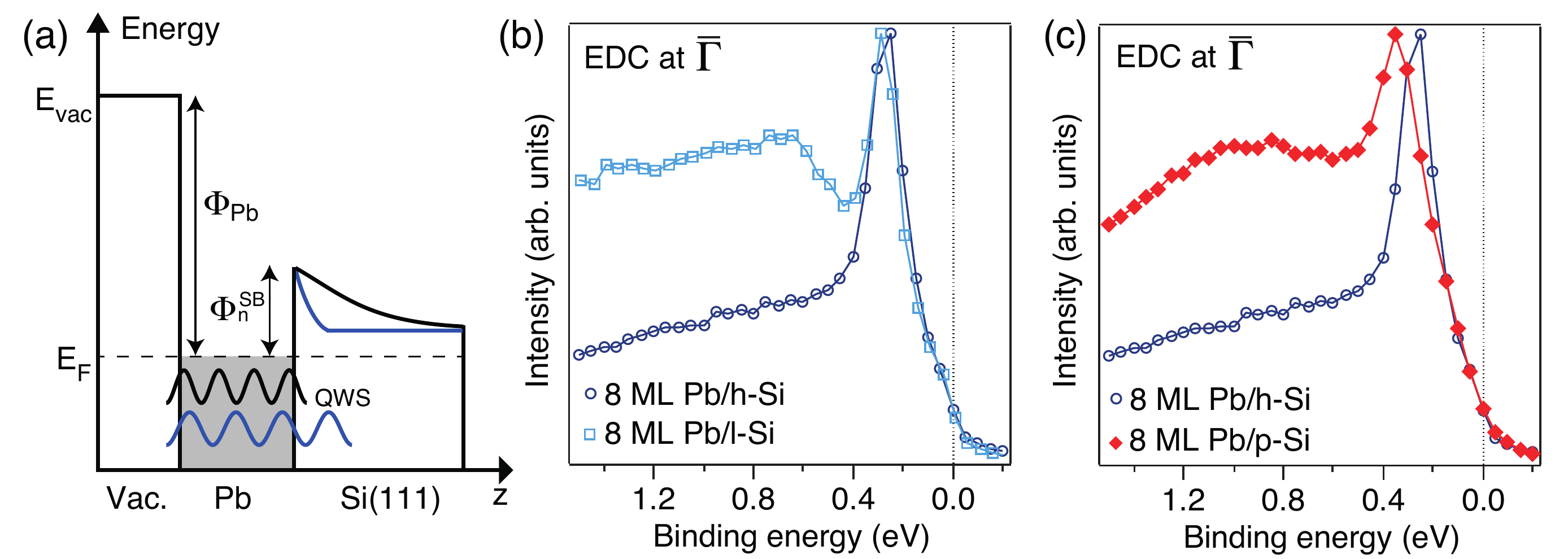}
\caption{(a) Schematic drawing of the confinement box of QWS in Pb/Si as a function of donor concentration. Comparison of normal emission spectra of a QWS in (b) Pb/h-Si and Pb/l-Si and in (c) Pb/h-Si and Pb/p-Si.}  
\label{Fig3}
\end{center}
\end{figure}

To summarize, while in Pb on n-type Si the Rashba parameter is found to increase with donor concentration, in Pb on p-type the opposite dependence is likely present, i.e. a decrease of $\alpha_{RB}$ with increasing acceptor concentration.
\subsection{Semiclassical penetration model}
In the following we discuss the results of the study of Ref. \cite{Zhang:2010} which investigated the influence of the barrier shape at the metal-substrate interface on the energetic positions of occupied and unoccupied QWS in Pb on n-Si(111) using scanning tunneling spectroscopy. The authors follow the idea that the Schottky barrier is independent on the doping concentration (corresponding to the Schottky limit) and that the confinement barrier for electrons approaching the substrate interface can be modeled via
\begin{eqnarray}
eV(z) = \left[(E_{[111]} + \Phi_n^{SB}) - \frac{eN_D}{2\epsilon_0\epsilon_r}(z-z_D)^2\right]\Theta(z-z_D)
\end{eqnarray}
where $E_{[111]}$ = 9.685 eV is the distance between the top of the s-like band of Pb in the $\Gamma$-L direction, $z_D$ is the depletion layer width and $\Theta$ is the Heaviside function. Based on this model one would expect a larger penetration depth of the envelope function of the QWS into the substrate with a higher doping concentration, as schematically illustrated in Fig. \ref{Fig3}(a). According to the particle-in-the box model the eigen-energy $E_n \propto d_{eff}^2$ of the state should then decrease because the effective width of the confinement box increases.
However in their measurements the scientists observe exactly the opposite behavior; the occupied levels in Pb on the heavily n-doped Si substrate are at larger energies compared with states in Pb on the lightly n-doped Si (compare data presented in Fig. 2 of Ref. \cite{Zhang:2010}). Because of the obvious conflict between the model and the experimental findings the authors concluded that the measured energetic positions of the QWS are an artifact of the measurement technique. However, in our experiment we find the same behavior of the energetics as a function of donor concentration which is exemplary shown in Fig. \ref{Fig3}(b). The state in Pb/h-Si is at lower binding energy compared to the state in Pb/l-Si(111), which strongly supports the interface dipole model for Schottky barrier formation. 
For clarity in Fig. \ref{Fig3}(c) we compare also normal emission spectra of QWS in Pb on the heavy p- and heavy n-type doped Si substrate. As already discussed the QWS in Pb/p-Si is more delocalized than in Pb/h-Si resulting in a larger binding energy at $k_{\parallel}$ = 0 \AA$^{-1}$.
\newline
\subsection{Electric field contribution to the Rashba effect}
In the originally derived Rashba model \cite{Bychkov:1984} an electric field transforms via the Lorentz transformation into an effective magnetic field, which causes the energy splitting between spin-up and spin-down bands of size $\Delta E = -\mathbf{\mu_s}|\bold{B_{eff}}|$, with the spin magnetic moment $\bold{\mu_s}$ = -$\frac{g_{e}\mu_{B}}{\hbar}|\bold{s}|$ and $\bold{B_{eff}} = -\frac{\hbar}{m_ec^2}(\mathbf{k_{\parallel}}\times \mathbf{E})$. From the donor density ($N_D$) and the built-in potential ($U_0$) we estimate the electric field at the metal-substrate interface via \cite{Brillson:1993}, 
\begin{eqnarray}
|\mathbf{E_{\text{M-S}}}| = \sqrt{\frac{2qN_{D}U_0}{\epsilon_{0}\epsilon_{r}}}
\end{eqnarray} 
and determine the corresponding energy splittings at $k_{\parallel}$ = 0.24 \AA$^{-1}$. For Pb/h-Si (high n-type doping) we obtain $E_{\text{M-S,h}} = 1.37$ MVm$^{-1}$ and $\Delta E_{h} = 2.4 \cdot 10^{-10}$ eV, for moderate doping $E_{\text{M-S,m}} = 0.36$ MVm$^{-1}$ and $\Delta E_{m} = 6.4 \cdot 10^{-11}$ eV, and for the low doping $E_{\text{M-S,l}} = 0.27$ MVm$^{-1}$ and $\Delta E_{l} = 4.8 \cdot 10^{-11}$ eV. The trend in the energy splittings as a function of doping reflects our experimental findings, but the calculated absolute values are eight orders of magnitude too small and cannot explain the difference of 14 meV in the measured spin splittings of Pb/h-Si(111) and Pb/l-Si(111). The results are summarized in Tab. \ref{Beff_table}. 
\begin{table}[htb]
\begin{tabular}[c]{|c|c|c|c|c|}
\hline
Doping & $N_D$ [cm$^{-3}$] & $E_{\text{M-S}}$ [MV/m] & $B_{\text{eff}}$ [T] & $\Delta E$ [eV] \\
\hline 
\hline 
low & $4.65 \cdot 10^{13}$ & 0.27 &  $8.4 \cdot 10^{-7}$ & $4.8 \cdot 10^{-11}$ \\
moderate & $8.38 \cdot 10^{13}$ & 0.36 &  $11.1 \cdot 10^{-7}$ & $6.4 \cdot 10^{-11}$ \\
high & $1.14 \cdot 10^{15}$ & 1.37 & $4.24 \cdot 10^{-6}$ & $2.4 \cdot 10^{-10}$ \\
\hline 
\hline
\end{tabular}
\caption{Donor density, interface electric field, effective B-field and energy splitting according to the Rashba model.}
\label{Beff_table}
\end{table}

Based on the bare Rashba model one would also expect the largest spin splitting in Pb QWS on p-type Si, because of the largest electric field due to the high acceptor concentration. However, this system feature actually the smallest spin splittings which is a further indication that the simplified original Rashba model does not hold for Pb QWS. 
\end{document}